# Helium abundance in giant planets and the local interstellar medium

L Ben-Jaffel[1,2] and I Abbes[1,2]

[1] UPMC Univ. Paris 06, UMR7095, Institut d'Astrophysique de Paris, F-75014, Paris, France
[2] CNRS, UMR7095, Institut d'Astrophysique de Paris, F-75014, Paris, France

E-mail: bjaffel@iap.fr

**Abstract**. The sun and giant planets are generally thought to have the same helium abundance as that in the solar nebula from which they were formed 4.6 billion years ago. In contrast, the interstellar medium reflects current galactic conditions. The departure of current abundances from the primordial and protosolar values may help trace the processes that drive the nucleosynthesis evolution of the galaxy and planetary interior formation and evolution. The Galileo probe measured the He abundance *in situ* the atmosphere of Jupiter, showing that He is only slightly depleted compared to the solar value. For Saturn, contradictory estimates from past Voyager observations make its He abundance very uncertain. Here, we use He 58.4 nm dayglow measured from the outer planets by the Voyager ultraviolet spectrometers to derive the He abundance in the atmosphere of Jupiter and Saturn. We also use the solar He 58.4 nm line measured by the Solar Heliospheric Observatory to derive the He abundance inside the focusing cone. Finally, we compare He abundances derived here with primordial and protosolar values, stressing the unique opportunity offered by inner heliosphere observations and future Voyager *in situ* local interstellar medium measurements to derive the He abundance in the very interstellar cloud in which we reside.

## 1. Introduction

According to the standard Big Bang Nuclueosynthesis (BBN) model, the universe was dense and hot at the beginning of its expansion phase. During four long minutes, nuclear reactions produced a significant fraction of the baryonic mass of the universe in the form of helium-4 (~25%) and traces of deuterium and lithium. At the end of the expansion period, the temperature of the universe dropped to a level that does not allow the cascade of nuclear reactions required to produce the other heavy elements of the periodic table. The mass embedded in those few initial species forms the so-called primordial baryonic mass. A few million years later, stars started forming with those primordial light elements (H, He, etc.). With the extreme density and temperature conditions in the interior of stars, nuclear reactions then produced new He and heavy species, which contaminated the interstellar medium through a variety of stellar mass loss processes.

One of the most intriguing results of the BBN model is that it allows one to predict a large primordial He mass fraction $Y_{BBN} = 0.2485 \pm 0.0002$ at the end of the expansion phase of the universe (based on the recent Planck determination of the baryon density) [1]. Because non-primordial He and heavy metals could only be produced in stars in contrast to D that is consumed, regions of poor metallicity could reflect the He condition not far removed from that which prevailed at the end of the expansion phase. This simple idea fueled the study of a collection of metal-poor extragalactic H II targets to obtain the He abundance versus the gas metallicity, usually expressed as the O I/H I ratio [2]. When extrapolated to zero metallicity, the technique should provide a measurement of the primordial He mass fraction $Y_p$ before the emergence of stars in the observed region. Such was the concept without taking into account systematic errors. Indeed, the application of this technique showed how the inferred $Y_p$ was sensitive to our level of knowledge of the driving processes in the gas phase of the studied targets. The continuous effort of the scientific community is illustrated in Figure 1 in terms of the temporal evolution of their results.

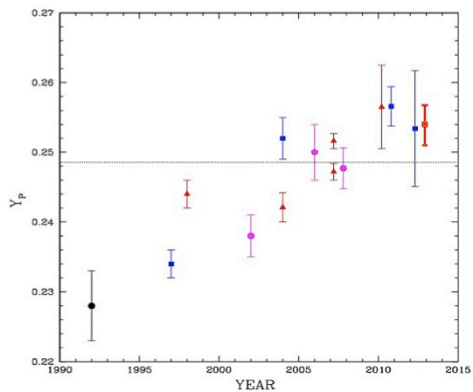

**Figure 1**. Primordial He mass fraction derived from extragalactic H II regions studies over time. Plot adapted from [1] with the newest value obtained by [2] added. For reference, the dotted line shows the $Y_{BBN}$ value.

Besides H II regions, the paradigm described above has been applied to the protosolar nebula, the formation and evolution of the sun, and the interstellar medium, which represent targets of high interest. For example, the comparison between the He mass fraction in the protosolar nebula and the photosphere of the sun revealed a diffusive settling that reduces the He and heavy species abundances relative to H I [3]. For giant planets, the Galileo probe *in situ* measurement also revealed a He mass fraction smaller in the outer layers of Jupiter compared to the protosolar value [4]. In the local interstellar medium, the ionized gas near early-B stars and in the Orient nebula indeed have a poor metallicity level, yet the exact He and metal abundances are also uncertain because gas depletion into dust grains complicates the data interpretation. In addition, the actual local interstellar medium (LISM) properties around the solar system may be different from the conditions that prevailed at the sun's birthplace in the galaxy because it may have migrated over its lifetime.

In the following, we revisit giant planets' He abundance using He 58.4 nm dayglow measured with Voyagers 1 and 2 ultraviolet spectrometers (UVS) during their encounters with the outer planets (Section 2). In Section 3, we also use the solar He 58.4 nm line but as measured by Solar Ultraviolet Measurements of Emitted Radiation (SUMER) instrument onboard the Solar Heliospheric Observatory (SoHO) to constrain the He abundance inside the gravitational focusing cone of the interstellar gas flow, which reflects the local interstellar cloud (LIC) composition. Finally, we compare our results to the protosolar and primordial He values in an attempt to connect heliospheric physics with the larger cosmological picture of the origin and evolution of the interstellar medium.

## 2  Giant planets helium abundance

An accurate measurement of the helium abundance in the atmosphere of giant planets is a key step in understanding the fundamental problem of the formation and evolution of giant planets in solar and extrasolar systems. In the past, using Voyagers 1 and 2 radio occultation observations to derive a pressure-temperature altitude profile, along with infrared emission spectra from most of the outer planets, a sophisticated approach was implemented to derive the He abundance by adjusting the atmospheric He /$H_2$ mixing ratio until the infrared spectra are best recovered [5]. However, the first *in situ* He mass fraction value $Y_{Galileo} = 0.234 \pm 0.005$ for Jupiter [4] provided by the Galileo Entry probe does not agree with the indirect Voyager value $Y = 0.18 \pm 0.04$ [5], casting doubt on the indirect method. Furthermore, contradictory estimates reported for the He abundance in Saturn using similar remote sensing methods with values ranging from $Y_{sat} = 0.06 \pm 0.05$ (pre-Galileo value [6]) to $Y_{sat} \sim 0.18$-$0.25$ (post-Galileo value [7]) added to the confusion and the lack of confidence the community has in the proposed values. In contrast to Jupiter, no space probe is actually planned to measure the Saturn He abundance *in situ*, thus making the diagnostic of this primordial parameter obtained with the Galileo probe incomplete.

*2.1 Voyager UVS observations & data analysis*

To derive the helium abundance, we use the first resonance line of He as emitted by the upper atmosphere of most planets at 58.4nm. The He 58.4nm dayglow thus far observed during the successive Voyager encounters with the outer planets corresponds to the solar He emission line back-scattered by the He atoms confined in their upper atmospheres (e.g., Table 1). The observed dayglow mainly depends on the He atmospheric total content, the eddy mixing power of the atmosphere below the homopause level, and the temperature profile above the homopause level.

The bulk of the atmosphere of the two giant planets is assumed to be composed mainly of $H_2$, He, and $CH_4$ (Figure 2). We solved the continuity equation for He and $CH_4$ as minor constituents, including molecular diffusion and eddy mixing parameterized by the eddy diffusion coefficient $K_H$. The background gas is assumed to be in hydrostatic equilibrium with a varying mean molecular mass.

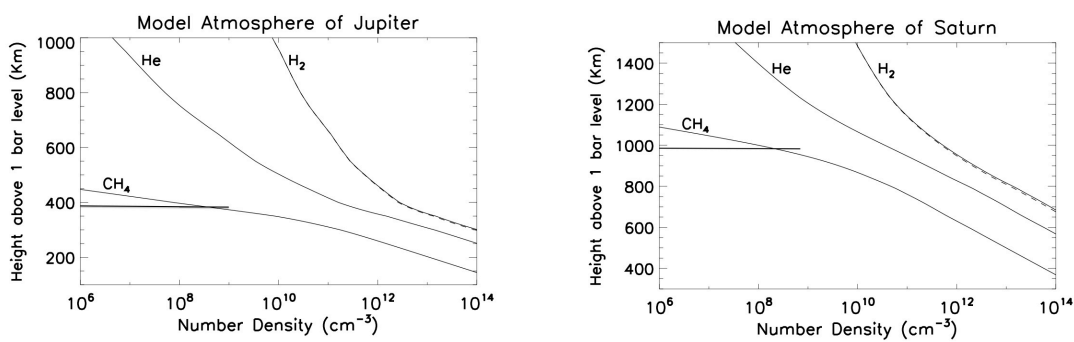

**Figure 2**. Atmospheric model derived in this study. A short horizontal line segment indicates the homopause level. (Left) Jupiter. (Right) Saturn.

To estimate a disk average range of $K_H$, we use stellar occultation observations made with the UVS instrument during spacecraft encounters with each of the planets. Of particular interest in our case is the light curve trend obtained for wavelength ranges that are sensitive to the absorption by $CH_4$. Because methane is the third-most abundant component of a giant planet's atmosphere, the corresponding light curve in the far ultraviolet shows a clear cut-off at the homopause level, offering the cleanest direct technique for locating its level and the corresponding eddy diffusion strength [8]. The $K_H$ values derived here (e.g., Table 1) are consistent with past estimates [8, 9].

The next parameter of importance for the He dayglow modeling is a good estimate of the temperature profile in the upper atmosphere, preferentially obtained for each planet during the same period of the airglow's observation. As shown in Figure 3, the different temperature profiles so far reported for Jupiter and Saturn do not show a noticeable variation (less than ~10%) around the homopause level [9, 10, 11]. The profiles shown in Figure 3 are explored as potential distributions of temperature versus altitude including estimated measurement and model uncertainties.

To derive the He abundance, we adjust the atmospheric helium mixing ratio $f_{He}$ until our radiation transfer model (RT) of the He 58.4nm sun-reflected emission line of the planet matches the corresponding observed brightness. The RT model includes photon absorption by molecular and atomic hydrogen and resonance scattering by He with partial frequency redistribution [12, 13].

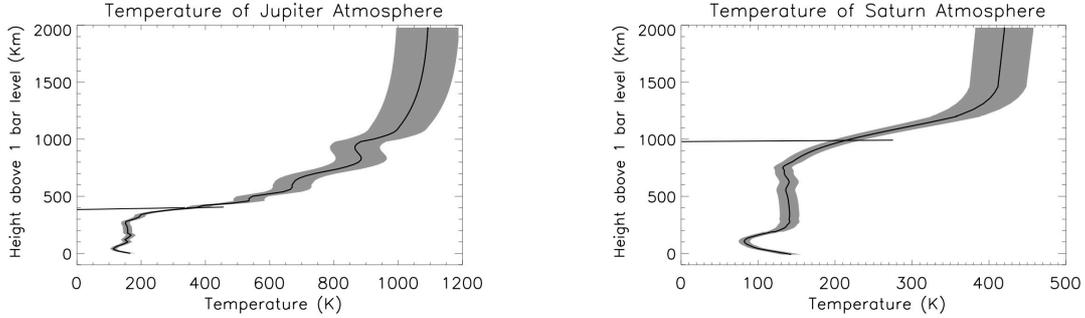

**Figure 3**. Temperature profile values (shaded) used in He 58.4nm Brightness RT model. The horizontal line segment indicates the homopause altitude level. (Left) Jupiter. (Right) Saturn.

Because the temperature depends on the mean molecular mass, its altitude profile and species distribution are updated iteratively for each mixing ratio ($f_{He}$) assumed deep in the atmosphere.

**Table 1.** Key input parameters for the He 58.4nm brightness radiation transfer calculations.

| Planet | Eddy diffusion $K_H$ (cm$^2$ s$^{-1}$) | 1 bar-level Temperature (°K) | He 58.4nm Brightness (Rayleigh) | He 58.4nm solar flux @ 1AU[a] (Photons cm$^{-2}$ s$^{-1}$) |
|---|---|---|---|---|
| Jupiter | $(2.5-4.0)\ 10^6$ | 165 | 6.5±1.1 (V1) & 5.7±1.2 (V2) | $3.832\ 10^9$ |
| Saturn | $(1.5-2.5)\ 10^7$ | 135 | 3.1±0.4 (V1) & 4.2±0.5 (V2) | $4.694\ 10^9$ |

[a] Integrated solar flux at the time of the V1 observation.

### 2.2 Helium Abundance in Jupiter & Saturn: results

The case of Jupiter is very interesting because the Galileo probe already measured the helium abundance with a high degree of accuracy, thus offering a key test for our He 58.4nm airglow-based technique. For the eddy diffusion coefficient range shown in Table 1 (from occultation observations), and for the temperature profiles shown in Figure 2, we derive a helium mixing ratio in the range $f_{He} \sim 0.159 \pm 0.034$ to fit both V1 and V2 UVS He 58.4nm airglow observations. The result corresponds to a helium mole fraction $q_{He} \sim 0.137 \pm 0.025$ and mass fraction $Y_{Jup} \sim 0.237 \pm 0.038$. This abundance is nicely consistent with the Galileo probe *in situ* values $q_{He} \sim 0.1359 \pm 0.0027$ and $Y_{Jup} \sim 0.234 \pm 0.005$ measured in the atmosphere of the planet [4]. The agreement between our result and the Galileo probe measurement successfully validates our technique.

The comparison of our model calculations with Saturn's He 58.4nm intensity from the Voyager UVS brightness observations shows a surprise. For the eddy diffusion coefficient values derived from occultation observations (e.g., Table 1) and the temperature profile shown in Figure 2 with the attached error bars, we derive $f_{He} \sim 0.151 \pm 0.025$ in the atmosphere of Saturn. This mixing ratio corresponds to a mole fraction $q_{He} \sim 0.131 \pm 0.02$ and mass fraction $Y_{Sat} \sim 0.226 \pm 0.03$, as summarized in Table 2. This is a higher abundance for Saturn than the value $Y_{sat} = 0.06 \pm 0.05$ found earlier by [6]. As a consequence, the mean molecular mass should be updated in the continuity equation [6]. In order to fit the background density with the helium mixing ratio derived here, the 1-bar level temperature in Saturn's atmosphere should be ~149 °K, a value larger than previously derived by [10]. Surprisingly, the high 1 bar level temperature supports the finding of [14], who suggested that a warmer atmospheric layer is required in their model of Saturn's formation.

**Table 2**. Helium abundance and temperature at the 1 bar level that best fit the He 58.4nm brightness observed by V1 & V2 for Jupiter & Saturn. Galileo probe results are shown for reference [4, 11].

| Planet | 1 bar-level Temperature (°K) | Mixing ratio ($f_{He}$=He /$H_2$) | He mole fraction ($q_{He}$) | He mass fraction ($Y_{planet}$) |
|---|---|---|---|---|
| Jupiter (Galileo) | 166 | $0.156 \pm 0.006$ | $0.136 \pm 0.003$ | $0.234 \pm 0.005$ |
| Jupiter (UVS) | 165 | $0.16 \pm 0.03$ | $0.137 \pm 0.025$ | $0.237 \pm 0.038$ |
| Saturn (UVS) | 149 | $0.15 \pm 0.025$ | $0.131 \pm 0.02$ | $0.226 \pm 0.030$ |

## 3  Local interstellar medium He abundance

The helium abundance in the local interstellar medium remains particularly uncertain because different LISM regions may have different metallicity and probably different helium abundances. Those differences are intimately related to the chemical evolution of the galaxy. To measure the LISM He abundance, three main regions have been considered in the past, namely the protosolar nebula, the Orient nebula, and the gas phase around early-B stars.

For the protosolar nebula, the sun's photosphere is usually used to derive the helium abundance. Using a sophisticated model of solar oscillation, the present-day helium abundance in the sun's photosphere is derived as $Y_{photosphere} \sim 0.2485$ [3]. From the photosphere He mass fraction, sophisticated 3D hydrodynamics and radiation transfer modeling is then used to derive the protosolar He and metals mass fractions that prevailed 4.6 Gy ago: $Y_{proto} \sim 0.2703$ and $Z_{proto} \sim 0.0142$ [15]. It is important to stress that the values are very model-dependent. Usually the difference between the photosphere and protosolar values is explained by diffusive settling that brings He and heavy species deeper toward the sun core. In addition, the protosolar nebula value represents an indication of the galactic composition at the time and birthplace of the sun, which may be different for other regions of the Galaxy and the actual epoch.

The second option to estimate the He mass fraction in the LISM is to use the bright and nearby Orient nebula. Traditionally, this nebula is considered as a standard source for the retrieval of the ionized gas chemical composition in the galactic environment around the sun. Using recombination and collision excited emission lines from ionized species, the conclusion was that the He mass fraction and the metallicity of the Orient nebula are slightly larger than in the sun's photosphere with $Y_{Orient} \sim 0.278$ and $Z_{Orient} \sim 0.0137$ [16]. However, both mass fractions are not independent of each other and hinge on the degree of success to properly include depletion into dust grains.

A third method used to derive He and heavy elements mass fractions in the LISM is to probe the ionized gas around early B-type stars [17]. Those short-lived stars have the interesting property not to have had time to migrate from their birthplace in contrast to old stars like the sun. In addition, the dust-grain phase seems nearly inexistent for those objects, which opens a good opportunity to derive a present-day cosmic chemical composition reference. Using few targets, [17] showed that the technique is very promising and derived $Y_{E-B-stars} \sim 0.276$ and $Z_{E-B-stars} \sim 0.014$ that are close to mass fractions derived from other techniques, yet sophisticated models are needed to reduce systematic uncertainties.

A common feature that we noticed for the three different targets considered above for the retrieval of cosmic abundances is that the ability to retrieve those abundances requires first a deep and sophisticated description of the physics of the region under study. Therefore, an accurate determination of He and metal mass fractions is closely related to our ability to model the physics in those astrophysical sites, much like the requirements in comparable studies for retrieving the primordial He mass fraction from extragalactic H II regions. In the following, we propose to apply the same paradigm to derive the helium abundance inside the local interstellar cloud (LIC).

*3.1    Helium abundance in the LIC*
The LIC can be described that the ISM cloud through which our solar system is actually moving. Either described as a separate objet or a piece of a much larger structure [18, 19], the pristine composition of the LIC remains a topic of vivid discussion [20, 21]. Following the same technique using the imprint of the LIC gas on nearby stellar emission lines, Extreme Ultraviolet Explorer (EUVE) spectra of six white dwarfs were used to derive H I and He I neutral density columns versus distance from the sun, with the interesting finding of a rather constant ratio H I/He I ~ 14 ± 2 ($2\sigma$ error bar) that could be used for the immediate neighborhood of the sun [22]. However, using the same EUVE instrument, another white dwarf (REJ 1032+532) showed a quite different ratio H I/He I ~ 7.2 [23], a finding that casts doubt on the uniformity of the H I/He I ratio initially derived when using only six target stars [22]. We believe the main problem resides in the short list of targets used that does not allow deriving reliable statistics.

Another tentative effort to derive the He abundance in the LIC was to use the He ionization degree versus distance from the sun, using local thermodynamic equilibrium (LTE) models applied on an extended set of seventeen hot white dwarfs observed by EUVE [24]. Besides the limitation in using LTE models for the selected white dwarfs, the large scattering that appears in the data makes the lack of gradient thus far reported for the He ionization degree (~40%) uncertain, particularly the extrapolation to short distances from the solar system. Here, it's important to remark that in contrast to many studies to derive the He abundance in different regions of the galaxy, the techniques thus far applied to the LIC make the assumption that the helium mass fraction is cosmic (He/H =0.1) and return partial information (ionization degree, H I/He I ratio, etc.). For reference, sophisticated modeling appeared in the literature, including radiation transfer effect for the photoionization modeling inside the LIC, and trying to make the link between the inner heliosphere *in situ* measurement and the distant LISM observations [25]. However, all those studies assume a cosmic value for the LISM He mass fraction and a constant H I/He I ratio in the LISM as derived from EUVE observations, assumptions that are questionable as discussed above.

Following the same line of thinking that prevailed in the retrieval of most cosmic chemical elements, we propose in the following a new and simple approach, principally based on in situ measurement obtained or planned for the near future, to derive an accurate and consistent He mass fraction in the LIC. This should ultimately make the link between heliospheric physics and cosmology in a rather elegant way.

*3.2    Helium abundance in LIC: back to the basics*
Because the solar system is moving inside the LIC, any sensor at any distance from the sun does provide in situ measurements. Over the last few decades, with space missions probing few but different neighborhoods around the sun, plasma, fields, and radiations in situ measurements became available over time [20, 21, 26]. Instead of going into a detailed description of available data, we prefer to list the ingredients required to derive the He abundance and assess their accuracy. Basically, the density of five species is required: neutral hydrogen (H I), neutral helium (He I), ionized hydrogen (H II), ionized helium (He II), and the double ionized helium (He III). In case one of the three ionized species is not available, on may include the electron number density Ne, using the property of plasma neutrality.

*3.2.1   Hydrogen & Helium neutral components in the LIC.* The interaction between the solar wind and the LIC produces a specific configuration with the heliopause interface that deflects all impinging ionized species but allow the neutral components to penetrate, yet suffering few charge exchanges on their way. Charge exchanges produce pick-up ions, which ultimately will produce a population of energetic neutral atoms. The important message here is that the interaction region between the solar wind and the LIC strongly modifies the incoming neutral population with the setup of many sub-population species (charge exchange history for neutrals, pick-up ions) that have specific kinetic properties but fully related to the primary LIC population. To access the undisturbed H I population, a sophisticated description of the interaction between the solar wind and the LIC flow is required including state-of-art MHD modeling for plasma and kinetic description for the neutrals [27, 28].

One of the pioneer technique to estimate the LIC neutral density is using the sky background glow produced by the solar radiation (H I Ly-α, Ly-β, He Ly-α) backscattered by interplanetary neutrals [29, 30]. The main strength of this local remote sensing technique is that the observed emissions are directly related to the unknown neutral population. However, opacity effects in a moving medium induce multiple scatterings that require state-of-the-art radiation transfer 3D models in addition to the 3D kinetic description of the H I neutrals themselves [31, 32, 33]. The technique provided valuable insights about the inner heliosphere, but unfortunately many problems persist to adequately describe the outer heliosphere sky background observations made by the Voyagers spacecraft in a consistent way (poorly constrained LIC parameters that are key inputs in all models, data calibration, missing atom's recoil during Ly-α scattering effect, etc.). As a consequence, the retrieved neutral hydrogen density remains poorly constrained (~ 0.1 - 0.25 cm$^{-3}$) despite an apparent consensus (figure 4).

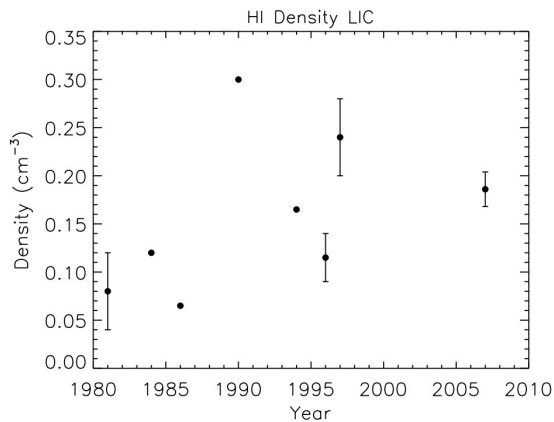

**Figure 4**. Neutral hydrogen number density estimated for the LIC by different techniques over time. Despite claims on its exact value, this quantity remains uncertain.

Nevertheless, the progress obtained by the community since the first discovery of the sky background glow is tremendous. It shows that 3D models should be improved to include most missing effects (see above), yet the data inversion should be implemented following sensitivity studies that scan all parameters of the problem (LIC & SW parameters, etc.). This technique requires expensive computer simulations [34], yet it should be one piece of the global approach proposed in this study.

A second technique that may give access to neutral number densities is to use pick-up ions *in situ* measurements made by several spacecraft inside the heliosphere (ACE, Messenger, Ulysses, Voyagers, etc.) [20]. Because pick-ions have the primary neutral population as their intrinsic source and are intimately related to the filtration processes at work near the heliopause, they do offer the opportunity to strongly constrain any model of interaction between the solar wind and the LIC plasma. Usually, obtained measurements have been efficiently used to derive the required neutral density level at the termination shock (TS) position [20]. Having the TS density level is a good step, yet sophisticated 3D plasma-neutral models are required to derive the undisturbed LIC neutral density. Because the phase space of unknown parameters is large (LIC & SW parameters), sensitivity studies are also required, yet within the frame of the global approach that includes both pick-ions and sky background radiation observations.

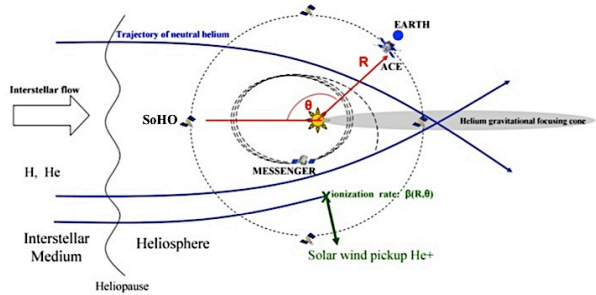

**Figure 5**. Sketch showing the He focusing cone and key space missions that provide *in situ* or remote observations of He atoms in the inner heliosphere. Reproduced from [35] where we added a reference to SoHO.

A third technique is fully related to the helium intrinsic distribution, particularly the formation of the focusing cone on the downstream of the flow (e.g., figure 5). Because it has a specific spatial distribution, the He cone was particularly used in the past to derive the LIC properties from different data sets (see [21] for a review). In that frame, during the two first years of the SoHO mission, we have observed the sun at the He resonance line (58.4nm) with the SUMER instrument along the orbit of SoHO in and outside the He focusing cone (figure 5). The aim was to detect the absorption feature that should appear in the solar He 58.4nm line profile due to the opacity of He atoms that fill the space between the sun and the spacecraft. In contrast to classical backscattering emission observations, this technique is free of solar flux and instrument calibration uncertainties [35]. Within the planned program, we were able to sample two times the angular region of the He cone twice, which allows checking and validating any spectral feature detection. In addition to the yearly repetition of the program, we also obtained time series of different quiet regions on the solar disk or off-disk in order to monitor any potential time variation of the source.

The SUMER instrument description and its performances before and in-flight are well described in [36]. We used the 1x300" slit pointing at either the quiet sun or off-disk from June 1996 to December 1997 in order to obtain time series of the He 58.4 line profile in the second order of the SUMER detectors. Because absorption by He atoms is almost negligible (less than 1 %) for the region outside the focusing cone, line profiles obtained around June of each year (when the SoHO orbital position is near the apex direction) are used to obtain a good reference for the undisturbed solar line profile. The selected reference line profile is obtained by iteration over available time series spectra in which instant line profiles do not depart more than a fraction of percent from the average line profile.

For the line profile inside the focusing cone, we applied the same statistical selection as for the undisturbed solar line profile but for time series obtained during the December period (SoHO in downwind direction). The comparison between the line profiles in- to out- of the focusing cone clearly shows a blue-shifted feature that corresponds to the expected absorption by the intervening He atoms in the cone zone (figure 6). We have checked and obtained similar results using the two available data sets recorded successively in 1996 & 1997.

To obtain the kinetic properties of the He atoms, we considered the so-called hot model, which includes solar gravity, radiation pressure, and photon and electron ionization [37]. To analyze the observed spectra, we used the solar line profile obtained outside the focusing cone as our reference and applied the He absorption provided by the hot model before comparing to the line profile observed inside the cone. Our first quick analysis shows a very extended absorption that cannot be fit with classical temperature levels which casts doubts on the assumptions made in the classical hot model.

This difficulty to fit observations inside the cone is not new as similar problems appeared for the interpretation of pick-up He ions distributions obtained by both ACE and MESSENGER spacecrafts [35]. Despite the large number of unknowns compared to available data, we could obtained the following LIC parameters N(He I) ~0.015 cm$^{-3}$ and V(He I) ~ 26.2±2 km s$^{-1}$, yet the temperature T(He I) remains poorly constrained. It is important to stress that the indicated values are uncertain and are still under investigation by incorporating ACE, MESSENGER, and Ulysses pick-up ion in situ measurements [35]. The interpretation of this kind of data is complex particularly for regions close to the sun [21], yet we believe the new SoHO/SUMER observations together with He pick-up ions measurements at different distances from the sun will provide new and strong constraints on both the He LIC parameters and the processes that define the He atoms distribution.

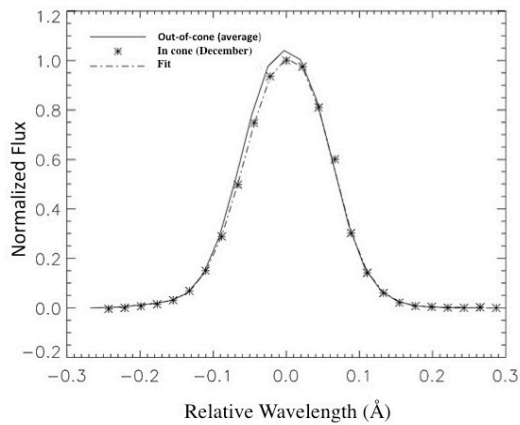

**Figure 6**. Solar He 58.4nm emission line observed by SoHO/SUMER in and out of the He focusing cone. The best fit to the in-cone line profile is shown (dash-dot line).

*3.2.2    Hydrogen, Helium, and electron plasma component in the LIC*. The heliosphere acts as a barrier that deflects most LISM electrons and ionized species. This means that we have no direct access to those species from inside the heliosphere. Fortunately, with the Voyager 1 & 2 getting outside the heliopause, we have a real and unique opportunity to measure charged components that could be accessed directly from plasma detectors, and combined radio waves and magnetometer measurements [26]. For the recent exceptionally low activity solar minimum, the heliopause position was close enough, which allowed Voyager 1 radio instrument to obtain the first indications on the LIC plasma density level (Ne ~ 0.05-0.08 cm$^{-3}$) [26]. With Voyager 1 & 2 operations to continue up to year ~2025, we have a rare chance to directly measure those key plasma parameters in the near future [38].

*3.2.3    How to derive the Helium abundance? Implementing the global approach*. In situ measurements outside the heliopause are fundamentally important for the retrieval of the helium abundance. Waiting for the exciting moment to see both Voyager 1 & 2 probes reach the LISM, a proper description from the inner boundary at the solar corona up to the outer boundary at the heliopause should be obtained using high-resolution MHD-kinetic models. This should be achieved following sensitivity studies that are expensive but efficient to capture the best space parameters that are consistent with available observations (plasma and pick-up ions *in situ* measurements, ENAs observations, Lyman-α data). Because the unknown parameters space is large, one should favor multiple-observations data analysis rather than focusing on a single dataset. This global approach showed its strength recently when Voyager 1 & 2 plasma measurements associated with the IBEX ribbon and a full sensitivity study were used to accurately constrain the LISM magnetic field [34].

## 4  Conclusions
We have reviewed the helium abundance at different sites of the Universe starting with the big bang era, then considered the protosolar nebula and giant planet composition, and finally the current local interstellar medium. From the studies thus far conducted on different astrophysical sites [1, 2, 3], the first lesson learned is that an accurate determination of the He abundance strongly constrains the key processes that control the formation of the astrophysical objects from the primordial Universe down to the current sun, and giant planets. The second important lesson is that the accuracy of the He abundance is strongly related the accuracy of the description of the physics of the object under study.

Here, we considered two targets—the giant planets of the solar system and the local interstellar cloud— to which we attempted to apply the initial steps of a global approach, following the path of prior studies conducted on extragalactic H II regions and the sun. For the giant planets, we propose a comparison of sophisticated radiation transfer models with the He 58.4nm airglow as observed during the encounters of the Voyagers spacecraft with the planets. In order to fit the planetary He 58.4nm airglow, our finding is that the helium mass fraction should be $Y_{Jup} \sim 0.237 \pm 0.038$ and $Y_{Sat} \sim 0.226 \pm 0.030$, respectively, for Jupiter and Saturn. Our results are consistent with the Galileo probe measurements for Jupiter, yet they stand in contrast with past findings for Saturn. For the higher He abundance obtained for Saturn, the stratospheric temperature should be raised up to ~149 °K in order to fit the background gas density thus far observed. With the helium abundance reported here, credible models become possible for the interior of Jupiter, Saturn, and extrasolar planets [39, 40].

For the local interstellar cloud, we found a situation with many observations and *in situ* measurements obtained by different techniques but lacking any strategy for linking the different pieces of the puzzle in order to obtain the He mass fraction. Here, we propose a new strategy based on a global approach. Starting from the inner boundary of the heliosphere at the position of the solar corona, we show how different data sets should be associated together to constrain the radial distribution of the He ionization rates, which should provide the number density of the He neutral component with a high accuracy. Also, *in situ* measurements by Voyager 1 & 2 plasma and/or radio emissions detectors in the near future, combined with magnetometer measurements, should provide the pristine density level of key ionized species in the LISM. When combined with sophisticated MHD-kinetic description of the heliosphere, those measurements should allow us to derive the density level of H I. The proposed approach should ultimately provide the helium abundance in the very interstellar cloud in which we reside. This would build a bridge between heliosphere physics and cosmology in a rather elegant way by showing that all past efforts so patiently made by the heliospheric community were not done merely for the narrow view of a specialized group, but rather as a piece in the big picture of the existence of the Universe and its evolution.

**Acknowledgments**
Authors acknowledge support from CNES, CNRS and UPMC. They thank the referee for very useful comments that helped clarify the manuscript.